\documentclass[aps,prl,twocolumn,floatfix,10pt]{revtex4-2}
\usepackage{graphicx}
\usepackage{color}
\usepackage[version=3]{mhchem}
\newcommand{\mymodA}[1]{\textcolor{black}{#1}}
\newcommand{\mymodB}[1]{\textcolor{black}{#1}}
\newcommand{\mymodC}[1]{\textcolor{black}{#1}} %red
\newcommand{\mymodD}[1]{\textcolor{black}{#1}} %blue
\begin{document}
\title{Length scales in electrolytes}
\author{Ioannis Skarmoutsos}
\affiliation{Laboratory of Physical Chemistry, Department of Chemistry, University of Ioannina, 45110 Ioannina, Greece}
%\email{iskarmoutsos@uoi.gr}
%
\author{Stefano Mossa$^*$}
\affiliation{Université Grenoble Alpes, CEA, IRIG-MEM-LSim, 38054 Grenoble, France}
%\email{stefano.mossa@cea.fr}
%
\date{\today}
\begin{abstract}
\noindent {\bf $^*$Author to whom correspondence should be addressed: stefano.mossa@cea.fr}
\vspace{.2cm}

\noindent The elusive presence of an anomalously increasing screening length at high ionic concentrations hampers a complete picture of interactions in electrolytes. Theories which extend the diluted Debye-H\"uckel framework to higher concentrations predict, in addition to the expected decreasing Debye length, an increasing significant scale of the order of at most a few ionic diameters. More recent surface force balance experiments with different materials succeeded in measuring increasing length scales which, however, turn out to extend over tenths or even hundreds of ionic diameters. While simulation work has managed to characterize the former, the latter still avoid detection, generating doubts about its true origin. Here we provide a step forward in the clarification of such a conundrum. We have studied by extensive Molecular Dynamics simulation the properties of a generic model of electrolyte, lithium tetrafluoroborate dissolved in ethylene-carbonate, in a vast range of salt concentrations continuously joining the Debye non-interacting limit to the opposite over-charged solvent-in-salt states. On one side, we have accurately determined the macroscopic concentration-induced structural, dielectric and transport modifications, on the other we have quantified the resulting nano-scale ions organization. Based only on the simulation data, without resorting to any uncontrolled hypothesis or phenomenological parameter, we identify a convincing candidate for the measured anomalously increasing length, whose origin has been possibly misinterpreted. 
\end{abstract}
\maketitle
In ionic fluids, the charge neutrality constraint implies the vanishing of the mean-field contribution to the system free energy, which is therefore fully determined by the electrostatic spatial correlations between the positive and negative ions~\cite{levin2002electrostatic}. The long-range nature of the Coulomb interactions induces properties which are absent in fluids composed of short-range interacting neutral particles, making the sophisticated theoretical tools created to deal with the latter difficult to adapt to the new scope~\cite{hansen2013theory}. Substantial part of the physics of electrostatic interactions is already included in the Debye-H\"uckel (DH) theory. In an electrolyte formed by monovalent ions dispersed in low concentrations in a solvent of dielectric permittivity $\epsilon$, the theory naturally establishes an exponential {\em screening} of the interactions between ions~\cite{hansen2013theory}, on distances dictated by the Debye length,
\begin{equation}
\lambda_D=\sqrt{\frac{k_B T \epsilon_o\epsilon}{e^2 c}}.
\label{eq:debye_length}
\end{equation}
Here $c=c_+ + c_-$ is the total ion concentration, $\epsilon_o$ is the vacuum permittivity, $T$ is the temperature, and $k_B$ the Boltzmann constant. At (very) low ionic concentrations, $\lambda_D$ is therefore predicted to {\em decrease} as $1/\sqrt{c}$.

At intermediate and high $c$, however, the picture complicates significantly, as theories assume that a spatial correlation function, $\Phi(r)$, related to mass or charge degrees of freedom, can be expressed in the alternative asymptotic forms,
\begin{equation}
\Phi(r)\propto
\begin{cases}
r^{-1} e^{-r/\lambda_\Phi} \\
r^{-1} e^{-r/\lambda_\Phi} \sin(2\pi r/d_\Phi+\psi_\Phi).
\end{cases}
\label{eq:Q(r)}
\end{equation}
Here, the first expression encodes an exponential (Yukawa) monotonic decay, recovering the DH theory in weak coupling conditions with $\lambda_\Phi=\lambda_D$. The second describes an oscillatory decay in the strong coupling limit, with $d_\Phi$ the period of the oscillations, and $\psi_\Phi$ a phase shift. Interestingly, now $\lambda_\Phi$ can be found to {\em increase} with $c$, assuming values as large as few ionic sizes, $a$, and always beyond $\lambda_D$ at the same $c$, a phenomenon called (regular) {\em underscreening}~\cite{rotenberg2018underscreening}. The two regimes are separated by the Kirkwood cross-over point, where $\lambda_\Phi\simeq a$.

The impact of intermediate and high ionic concentrations on the transport properties in general, and the features of the screening mechanism in particular, turns out to be a crucial element in applications, ranging from batteries technology~\cite{giffin2022role} to the development of bio-electronic sensors~\cite{kesler2020going}. Theory, often based on over-simplified models, has hence been increasingly complemented by extensive experimental work to verify the validity of the above predictions in realistic conditions. 
\begin{figure*}[t]
\centering
\includegraphics[width=0.99\textwidth]{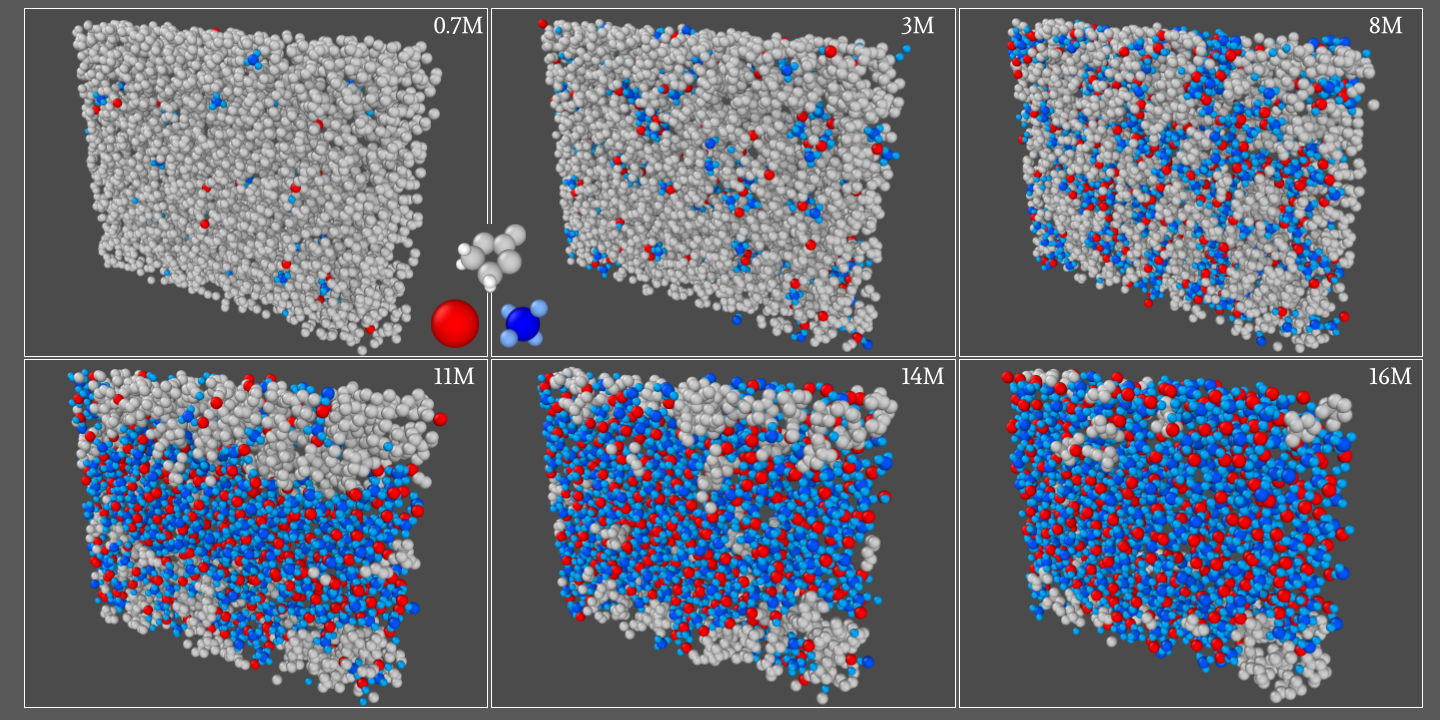}
\caption{Typical system configurations generated by MD simulation. In center left position, we show the chemical moieties composing the investigated ternary mixture, the \ce{Li+} cation (red), the \ce{BF4-} anion (blue), and the ethylene-carbonate solvent molecules (gray). In the six main panels, we show representative snapshots of the investigated system, at the indicated values of the salt concentration, $c$. We display system slabs of width $12$~\AA, oriented along one diagonal of the simulation box. On increasing $c$, the initially isolated ionic clusters (already at the edge of the DH limit for the present model) aggregate in increasingly larger ionic domains (red/blue clusters), developing in a unique percolated gel-like structure \mymodC{and, eventually, undergoing phase separation}. These modifications are discussed at length in the main text. 
}
\label{fig:snapshots}
\end{figure*}

Among the employed techniques, we mention here the Surface Force Balance (SFB)~\cite{hayler2024surface}, which measures directly the interaction force of two surfaces separated by the fluid sample. SFB is based on the Derjaguin-Landau-Verwey-Overbeek (DLVO)~\cite{smith2020forces,barrat2003basic} view that the overall interaction force between the two surfaces can be written as $F=F_{vdW}+F_{dl}$, where the first term accounts for the van der Waals forces, while the second represents the double layer interactions and can be expressed in the DH limit as $F_{\text{dl}}=4\pi\epsilon_o\epsilon(R_\text{s} / \xi)\phi_s^2e^{-d/\xi}$~\cite{smith2020forces}. Here, $R_s$ and $\phi_s$ are related to the geometry and the potential of the surfaces, respectively, $d$ is the distance between the surfaces, and the {\em bulk} electrolyte is described by $\epsilon$ and the (unknown) screening length $\xi$. 

\mymodB{Notable work~\cite{gebbie2013ionic,perkin2013stern,smith2016electrostatic,lee2017underscreening}} found that, indeed, at low $c$ the force between the two surfaces decays exponentially with $d$ and $\xi\simeq \lambda_D$, confirming the relevance of a surface-related measure to probe a bulk feature. Further measurements, though, conducted beyond the Kirkwood point on a range of materials (including inorganic salts in water, pure ionic liquids, mixtures of ionic liquids with organic solvents) provided an unattended observation. In addition to an oscillatory force at small $d$ (second expression of Eq.~(\ref{eq:Q(r)})) an exponentially decaying force exists at distances beyond the oscillatory part of the interaction, with a decay length increasing with $c$~\cite{smith2016electrostatic}. The observed underscreening, however, is {\em anomalous}\cite{jager2023screening,elliott2024known}, as the detected screening length reaches values as large as tenths to hundred (for ionic liquids) of $a$, in contrast with the predicted regular underscreening. (Rare negative results have also been reported as, for instance, in the Atomic Force Spectroscopy work of~\cite{kumar2022absence}.) 

To make things even trickier, a large corpus of computer simulation works of realistic models in different settings~\cite{keblinski2000molecular,coles2019correlation,zeman2020bulk,krucker2021underscreening,zeman2021ionic,jones2021bayesian,pivnic2019structural}, while unsurprisingly identifying regular underscreening, was unable to detect any sign of the anomalous form, generating doubts about both the true origin of the latter and the interpretation of the experimental data~\cite{jager2023screening}. (We note the remarkable exception of the restricted primitive model of ~\cite{hartel2023anomalous}, although in conditions not met in experiments~\cite{jager2023screening}.)

Here, we move forward on these issues, simulating by classical Molecular Dynamics (MD) methods a generic model of electrolyte, lithium tetrafluoroborate, \ce{Li+BF4-}, dissolved in ethylene-carbonate, \ce{EC}, in a vast range of concentrations, systematically investigating decisive structural, dielectric, and transport properties. Based on the simulation data only, without introducing any unverified assumptions or empirical parameters, \mymodB{and building on all previous work on the matter}, we propose a comprehensive explanation for the observed anomalously growing length, which might have been misunderstood in earlier analyses.
\begin{figure*}[t]
\centering
\includegraphics[width=0.85\textwidth]{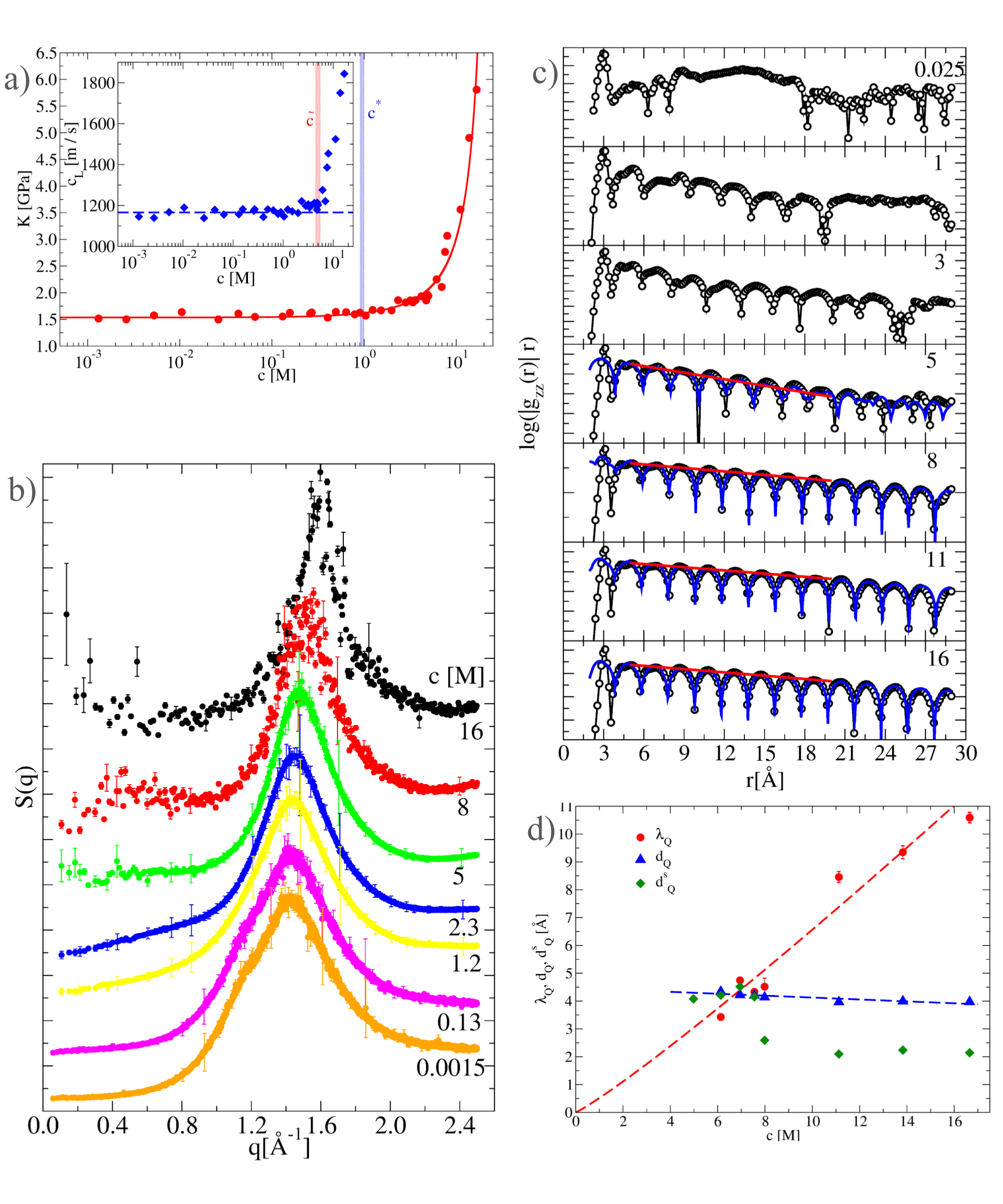}
\vspace{-1cm}
\caption{Modification with concentration $c$, of thermodynamic and structural properties. {\em a)} Variation of the bulk modulus, $K$ (closed symbols), together with the critical prediction   $\Tilde{K}\propto |c-c_{cr}|^{-\nu}$ (solid line), discussed in the main text. The vertical shaded blue line indicate the concentration $c^*\simeq$~1~M, whose relevance will be further discussed in the following. In the inset of the same figure we show the corresponding longitudinal sound velocity, $c_L(c)$, which shows an abrupt modification at $\tilde{c}=$~5~M (shaded red line). {\em b)} The static structure factors, S(q), at the indicated values of concentration. The important modifications visible at small wave vectors, $q$, are discussed at length in the main text. {\em c)} The  spatial charge-distribution functions, $Q(r)$, defined in the main text, at the indicated values of $c$ (solid symbols). At the highest $c$ we also show (blue and red solid lines) the results of a fitting procedure based on the oscillating exponential decay of Eq.~(\ref{eq:Q(r)}). {\em d)} Relevant length scales extracted by the fitting procedure. The red points clearly indicate the presence of regular overscreening, as detailed in the main text.}
\label{fig:modulus_and_s_q}
\end{figure*}
\section*{Results and discussion}
\label{sect:results}
\mymodB{Salient facts about our computations are given in the Methods Section and in the supplementary material.}
%, while complete detailed information together with a description of all raw data can be found in~\cite{skarmoutsos2025}.}  
We have employed LAMMPS~\cite{plimpton2007lammps} to simulate by Molecular Dynamics in the $(\text {NPT})$-ensemble at $T=$~400 K and ambient pressure a non-polarizable model for $\ce{Li+}\ce{BF4^-}$ dissolved in $\ce{EC}$~\cite{soetens1998molecular,masia2004ethylene,skarmoutsos2015li+}, in a vast range of salt concentrations, $c\in[0.0015:16.5]$~M. 
Typical system sizes range from $N=$~7$\times$10$^3$ to 10$^5$ at high and low values of $c$, respectively, while we considered a largely varying number of ion pairs, $N_{pairs}$ (see supplementary material). \mymodC{(Note that our model at the investigated temperature stays in a single-fluid phase up to concentrations exceeding $5$~M, with phase separation occurring at even larger values of $c$ (see below). Additional data, shown in the supplementary material, show that at $T=$~300 K phase separation occurs already at $c\simeq 3$~M, consistent with experimental measurements~\cite{allen2012solvent}.)}

In Fig.~\ref{fig:snapshots} we show the components of the considered ternary mixture, together with a few representative system snapshots, at the indicated values of the concentration. (We have used OVITO~\cite{stukowski2009visualization} for visualization.) Particular care has been taken to ensure proper thermodynamic equilibrium: each one of the investigated systems, at different values of $c$, has been prepared independently and subsequently aged on time scales of the order of at least 50~ns. System coordinates extracted from the following 30~ns production runs have been collected, and employed to compute the  necessary observables. Additional out-of-equilibrium runs in the presence of external electric fields of different magnitudes have also been performed (see below). \mymodD{We have also verified that no relevant issues due the finite sizes of the investigated systems or the MD algorithms used for propagating the trajectories is present, as discussed in the supplementary material.} 

{\bf Elastic behavior.} By increasing $c$, the volume modifications consequent to the (NPT)-ensemble evolution induce an overall density variation as large as about 50~$\%$ \mymodB{(see Fig.~S1 in the supplementary material)}. Changes related to the charge loading are clear from Fig.~\ref{fig:modulus_and_s_q}. In the main panel of Fig.~\ref{fig:modulus_and_s_q}~a) we show the bulk modulus determined from the volume fluctuations as $K=\beta_T^{-1}$, where $\beta_T=(\langle V^2 \rangle-\langle V \rangle^2)/(\langle V \rangle k_B T)$ is the isothermal compressibility. (We indicate with $\langle\rangle$ the ensemble average.) $K$ maintains the (pure) \ce{EC} value $\simeq$~1.5~GPa for $c<c^*\simeq$~1~M (vertical shaded blue line), where it starts a very steep increase with concentration. 
%(An analogous increase is found for the constant pressure specific heat, $C_P$, not shown~\cite{skarmoutsos2025}.) 
In the same plot we also show (solid line) the function $\Tilde{K}\propto |c-c_{cr}|^{-\nu}$ with $\nu=0.88$ and $c_{cr}=20.8$~M, which satisfactorily describes the data, and whose rationale we will discuss below. Note that we have refrained from a regular fitting procedure, and have instead fixed the value of $\nu$~\cite{stauffer2018introduction} and optimized $c_{cr}$ and the prefactor. These results indicate important modifications of the mechanical response of the material with $c$, which is even more evident from the abrupt variation of the longitudinal sound velocity, $c_L=(K/\rho)^{1/2}$ ($\rho$ are the mass densities), at $\tilde{c}\simeq$~5~M (shaded red line in Fig.~\ref{fig:modulus_and_s_q}~\mymodB{b)}). \mymodA{(Unfortunately, we have not been able to find any data in the literature referring to materials similar to those investigated here, neither concerning the concentration dependence of the bulk modulus, nor of the sound velocity. Additional work is needed in this direction.)}

{\bf Static structure factor.} Modifications in the mechanical response are accompanied by important changes in the structural features. In Fig.~\ref{fig:modulus_and_s_q}~c) we show the static structure factor, $S(q)=1/N|\sum_l b_l \exp{(i\mathbf{q}\cdot\mathbf{r}_l})|^2$, at the indicated values of the concentration. Here, $N$ is the total number of atoms, $\mathbf{r}_l$ are the atomic coordinates, $\mathbf{q}$ is the wave vector, and $b_l$ are the neutron scattering lengths. At very low concentration (bottom), the $S(q)$ is typical of a regular molecular liquid, with a first diffraction peak around $\simeq$~1.5\,\AA$^{-1}$ and a slight shoulder at $\simeq$~1.2\AA$^{-1}$. \mymodC{Interestingly, no particular modifications are visible even for concentrations slightly exceeding $c\simeq$~5 M, indicating that, at $T=$~400 K the regular fluid phase is stable in a quite large $c$-range.}

On increasing $c$, additional structure starts to be visible at low-$q$ values, which develops in a fully developed pre-peak at $\simeq$~0.4\,\AA$^{-1}$. This feature corresponds to long-range fluctuations on a length scale of the order $2 \pi / q^*\simeq$~16\,\AA, often observed in ionic liquids (see, among many others,~\cite{triolo2007nanoscale,triolo2021liquid}) and associated to the mesoscale organization of the three-dimensional network formed by polar aggregates~\cite{araque2015modern, annapureddy2010origin}. At the highest available concentration, $S(q)$ eventually diverges for $q \rightarrow 0$, signaling the insurgence of a phase separation between solvent-rich and highly charged (anions {\em and} cations) domains (we have discussed this point in ~\cite{mossa2018re}, although in a different context). Note that the formation of these ionic structures stiffened by very strong local attractive interactions is probably the responsible of the increase of the bulk modulus of Fig.~\ref{fig:modulus_and_s_q}~a). \mymodC{Altogether, these data indicate that our runs encompass the entire interesting $c$-range, spanning from the completely diluted limit, through an extended region where an homogeneous ionic fluid can be observed at equilibrium, to the opposite completely phase separated solvent-in-salt condition.}

{\bf Pair distribution functions.} Insight on the structural ionic organization can be further strengthened in the real-space, where we are guided by Eq.~(\ref{eq:Q(r)}). \mymodB{For our symmetric binary salt with charges $\pm e$, the relevant spatial charge-distribution function can be expressed as $g_{zz}(r)=g_{++}(r)+g_{--}(r)-2g_{+-}(r)$~\cite{coles2019correlation,hansen2013theory}, where the $g_{\alpha\beta}$ ($\alpha,\beta=\pm$) are the pair distribution functions pertaining to the cations and anions, respectively, and $r$ is the distance between the centers of mass of the involved ions}. We show a subset of our data in Fig.~\ref{fig:modulus_and_s_q}~c), in the form $Q(r)=\ln(r\, |g_{zz}(r)|)$, at the indicated values of $c$. The modifications  are notable, with a main peak at the $c$-independent position $r_M\simeq 3$~\AA, followed at low-$c$ by a quite unstructured pattern at larger distances. Some ionic nanostructuration only appears at $c\simeq c^*$, while a plainly developed oscillatory screened decay of the form of Eq.~(\ref{eq:Q(r)}) takes shape at $c\simeq \tilde{c}$~\cite{gavish2018solvent}. (We note that $c^* $ and $\tilde{c}$ partition the investigated $c$-range in three regions which reassuringly correspond to the domains of the phase diagram of Fig.~1 in~\cite{gavish2018solvent}, for a mean-field Poisson-Nernst-Planck model of electrolyte.) 

We have fitted the data with the sum of two oscillating decaying exponentials of Eq.~(\ref{eq:Q(r)}) (blue lines)\cite{zeman2021ionic}, while we plot in red the exponential term only, corresponding to the {\em largest} $\lambda_Q$. \mymodB{(At the lowest $c$, properly calculating the $g_{zz}$ is non-trivial, and implies considering extremely large system sizes~\cite{zeman2020bulk,zeman2021ionic}, a task beyond the scope of the present work, and we have therefore refrained from trying fitting those data. At higher $c$ the sum of two terms was needed to account for the presence of different mechanisms of different relative strengths, as it has been observed in previous work~\cite{zeman2021ionic}.)} The determined fitting parameters, $\lambda_Q$ and $d_Q$ are shown in Fig.~\ref{fig:modulus_and_s_q}~d). Regular underscreening is clear, as already detected in~\cite{keblinski2000molecular,coles2019correlation,zeman2020bulk,zeman2021ionic,krucker2021underscreening,jager2023screening}, while the apparent oscillation period very mildly decreases with $c$. \mymodB{Note that at around $c\simeq$~7.5~M, the length scales both match the value of the mean ion diameter, ($a\simeq$~4.5 \AA, see below), and for $c\lesssim\tilde{c}$ a clear determination of the two becomes excessively difficult.}

We conclude our discussion on this point by plotting in the same figure (diamonds) the spatial period of oscillation associated to the {\em shortest} decay length, $d^s_Q$. Interestingly, $d^s_Q$ decreases of a factor of two in the region $c>\tilde{c}$, a finding already discussed in~\cite{coles2019correlation,coupette2018screening} (and references therein). On the basis of those works,  $\tilde{c}$ should therefore correspond to the transition from an oscillatory charged-dominated to an oscillatory density-dominated decay of correlations, at the Fisher-Widom line. We will put in perspective this observation, together with the Kirkwood crossover at lower $c$, in the following.
\begin{figure}[t]
\centering
\includegraphics[width=0.49\textwidth]{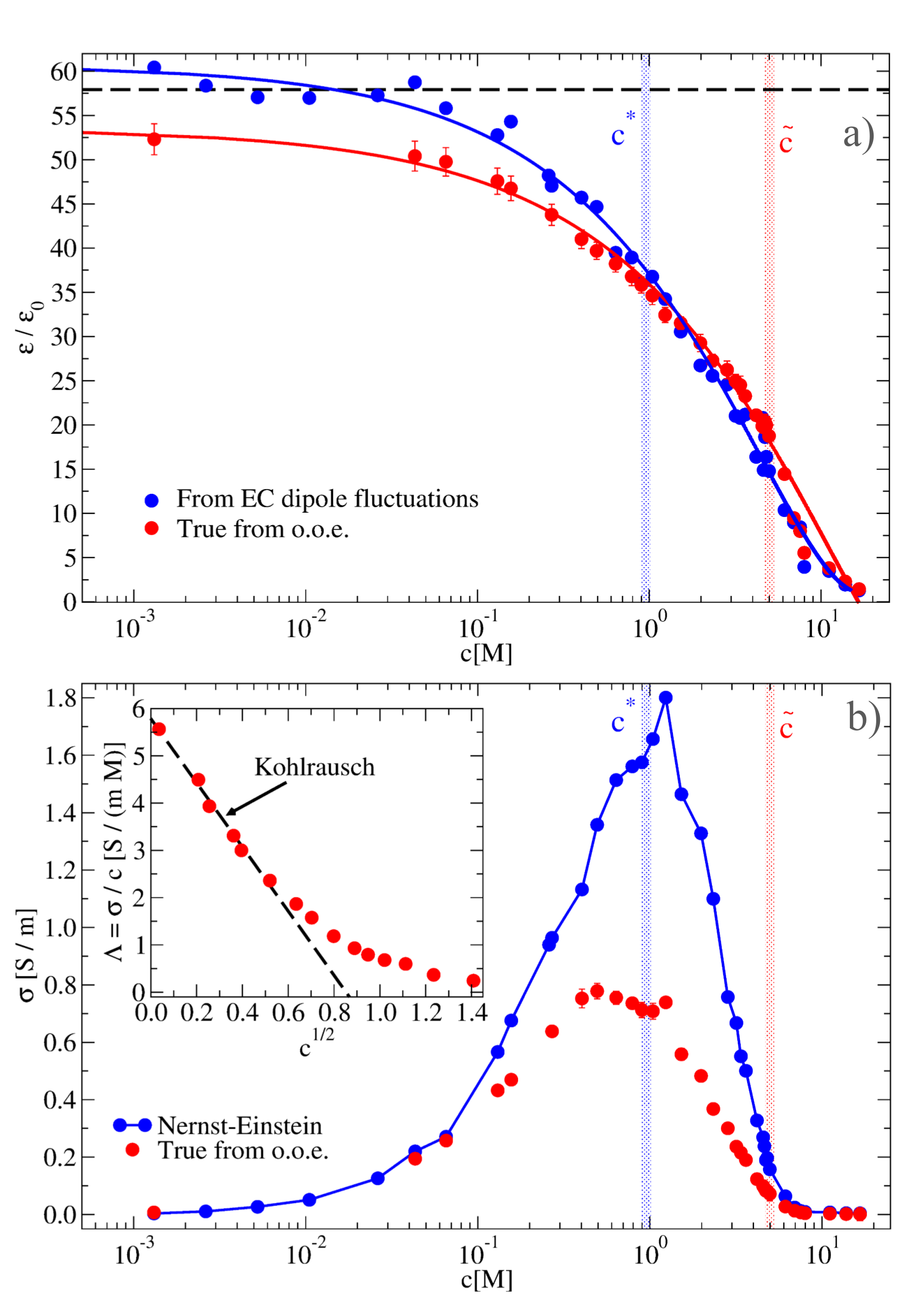}
\caption{Concentration dependence of the dielectric permittivity and the ionic conductivity. {\em a)} The dielectric permittivity $\epsilon(c)/\epsilon_o$, determined both from the \ce{EC} dipole fluctuations (blue circles) and via the out-of-equilibrium technique (red circles) detailed in the main text. The data are well represented by the fits indicated by the solid lines and described in the main text. The horizontal line indicates the experimental value of $\epsilon$ for pure EC at $T=400$~K~\cite{you2015dielectric}. The vertical shaded lines indicate $c^*$ (blue) and $\tilde{c}$ (red). {\em b)} The ionic conductivity, $\sigma(c)$ determined both from the ions diffusion coefficients in terms of the Nernst-Einstein approximation (blue) and via the out-of-equilibrium technique (red circles) detailed in the main text. In the inset of the same figure we show the same out-of-equilibrium data as the molar ionic conductivity, as a function of $\sqrt{c}$, to emphasize the Kolraush-like behavior at low $c$. These data are described in details in the main text.}
\label{fig:epsilon_and_sigma}
\end{figure}

{\bf Dielectric response and conductivity.} The above mechanical and structural modifications obviously influence both dielectric and ionic transport features. We show in Fig.~\ref{fig:epsilon_and_sigma}~a) and~b) the dielectric permittivity, $\epsilon$, and the ionic conductivity, $\sigma$, which will be important in the following. We have calculated $\epsilon(c)$ following two distinct routes. First, $\epsilon$ was determined from the fluctuations of the total simulation box dipole moment, $\mathbf{M}$, as $\epsilon/\epsilon_o=\epsilon_S/\epsilon_o=1+(\langle M^2\rangle-\langle M\rangle^2)/(3Vk_B T\epsilon_o)$. Note that this corresponds to the solvent ($S$) dipole moment contributions, those associated to the salt vanishing for symmetry. \mymodB{(The experimental value for pure $\ce{EC}$ at $T=400$~K is $\epsilon_S/\epsilon_o\simeq 59\pm 6$~\cite{you2015dielectric}.) We show the data in Fig.~\ref{fig:epsilon_and_sigma}~a).}

In addition, we have considered the out-of-equilibrium route of the finite field formalism of~\cite{cox2019finite}, which correctly accounts for the collective correlations of the ions. At each value of $c$ we have executed 4 runs under the presence of an external electric field $\mathbf{E}=E\mathbf{\hat{x}}$, collecting the solvent average polarization $\langle P^S_x\rangle$. We have next extracted $\epsilon$ by fitting $\langle P^S_x\rangle$ vs $E$, as described in~\cite{cox2019finite}. At $T=$~400~K in the diluted limit we find $\epsilon/\epsilon_o\simeq$~54.3, in good agreement with experimental findings. Both sets of data decrease with $c$, the dielectric decrement already discussed, among others, in~\cite{sega2015kinetic,gavish2016dependence}. We show by solid lines fits in the form $\epsilon(c)=\epsilon_o-A\,c+B\,c^{3/2}$ (\cite{cox2019finite} and references therein), which works quite well in both cases in the entire $c$-range. It is interesting to note that already a vanishing salt concentration significantly suppresses the dielectric response and $\epsilon<\epsilon_S$ up to $c\simeq c^*$, where the curves cross turning things around, in the limit of our numerical accuracy. We also notice the very low value reached at the highest $c$.

The ionic conductivity, $\sigma(c)$, is shown in Fig.~\ref{fig:epsilon_and_sigma}~b). We have, again, followed two independent routes. First, a non-interacting value can be extracted from the ions diffusion coefficients, $D_\pm$, via the Nernst-Einstein relation $\sigma_{\text{NE}}=e^2/V k_B T(N_+ D_+ + N_- D_-)$ (blue circles). We have also determined the true (collective) value via the finite field formalism of~\cite{cox2019finite}. Under the same external field conditions described above, we have first determined the time evolution of the ionic polarization, $P^{\text{ions}}_x(t)$, and calculated the average ionic current density as $\langle j_x\rangle=\partial P^{\text{ions}}_x(t)/\partial t$. The conductivity can next be comfortably determined from a linear fit of $\langle j_x\rangle$ {\em vs} $E$.

The data show a picture consistent with the known general behavior. In the inset we show the molar true ionic conductivity, together with the expected Kolrausch dependence, $\sigma/c\propto c^{1/2}$, at small $c$. At low concentrations in the Debye-H\"uckel regime up to $c_o\simeq$~0.1~M, the two sets of data superimpose, as expected. Next, they follow quantitatively the same behavior, reaching a maximum at $\simeq c^*$ and eventually decreasing to extremely low values at $\tilde{c}\simeq$~5~M. Note that, at the maximum, $\sigma_{\text{NE}}/\sigma\simeq$~2.25, stressing the importance of ionic correlations already at concentration less than 1~M. 

{\bf The (local) \ce{Li+} coordination shell.} We now build a {\em microscopic} view of the ions structural organization by investigating the features of the local environment of $\ce{Li+}$, i.~e., elucidating the modification with the salt concentration of the composition of the coordination sphere, $\mathcal{C}$. The coordination features of \ce{Li+} in ethylene-carbonate in the limit of infinite dilution are well known, with a tetrahedral structure which we also discussed in~\cite{skarmoutsos2015li+}. The effect of the anion is still far to be completely clarified, see the recent~\cite{vonwaldcresce2015anion} or our~\cite{skarmoutsos2015li+,ponnuchamy2018solvent,jiang2016anion}.
\begin{figure}[t]
\centering
\includegraphics[width=0.49\textwidth]{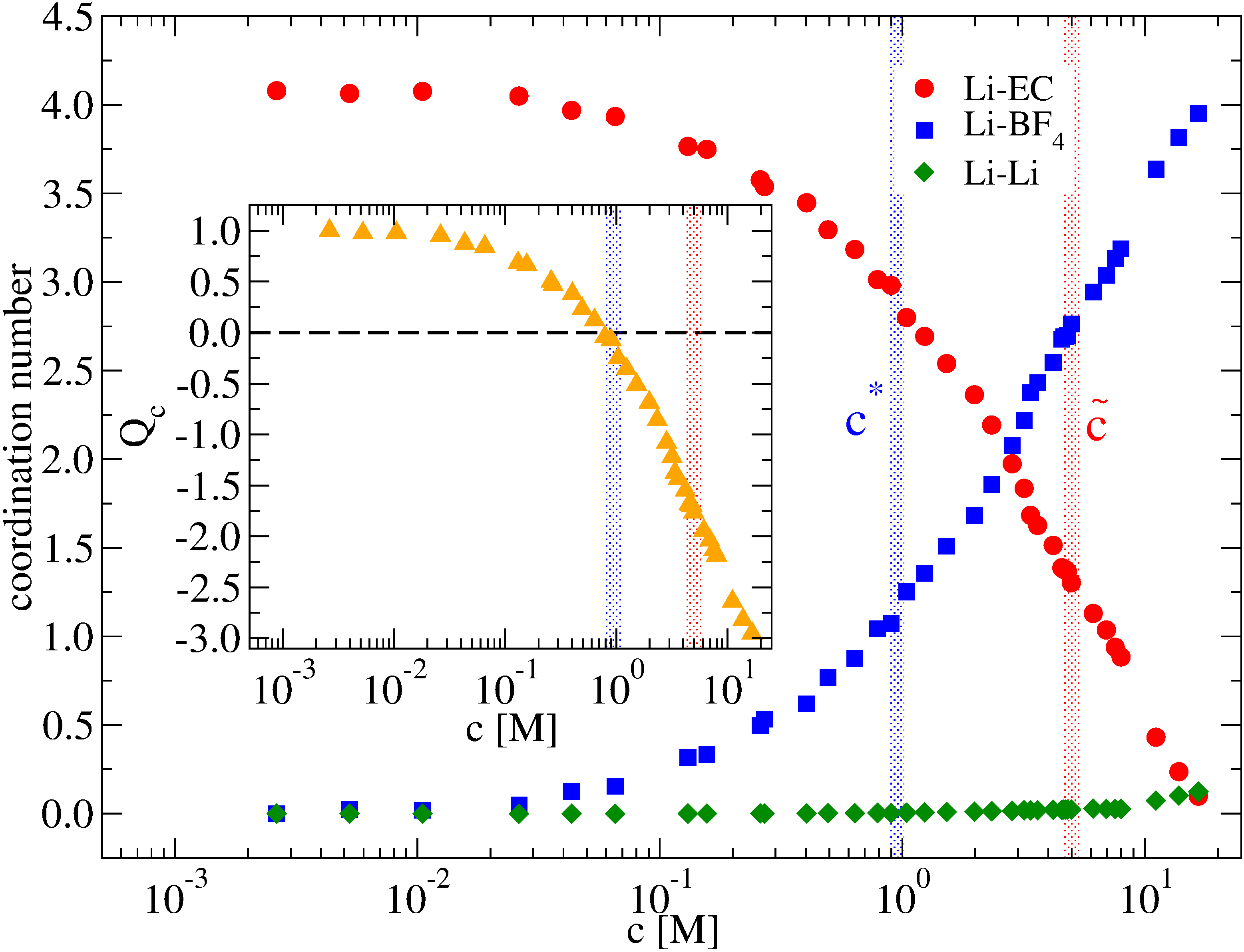}
\caption{Local coordination properties of the \ce{Li+} ion. {\em Main panel:} Concentration dependence of the coordination numbers of the indicated chemical species with the cation. The first coordination shell is formed by 4 molecules in all conditions, the number of coordinated anions steadily increasing with $c$. At around $\simeq$~3~M, the shell is composed, on average, by two solvent molecules and the same number of counter-ions. The number of coordinating cations (green) is negligible in all conditions. {\em Inset: } Based on the data of the main panel, we have determined the total charge, $Q_c$, of the coordinated (dressed) \ce{Li}-ion, which steadily decreases from $Q_c$=~1 in the diluted limit and cancels at $c^*$, while exactly 3 anions coordinate at $\tilde{c}$. All data are discussed at length in the main text.} 
\label{fig:coordination_Li}
\end{figure}

\mymodB{We have proceeded by counting the number of solvent molecules and co- (counter-) ions contained in spheres centered on $\ce{Li+}$ and of radius $r_1=2.6$~\AA\; and $r_2=3.8$~\AA\,, respectively, as extracted from the position of the first minimum in the partial pair distribution functions (see Fig.~S4 in the supplementary material).} The results are shown in Fig.~\ref{fig:coordination_Li}. Although the total coordination number is~4 in the entire investigated $c$-range (most probably due to employed non-polarizable force field), the number of $\ce{EC}$ molecules and $\ce{BF4-}$ is modified by $c$. Indeed, starting from the highly diluted state where the entire coordination sphere is composed by solvent molecules, by increasing $c$ the number of $\ce{EC}$ molecules decreases first slowly, for $c<c_o\simeq$~0.1~M, next faster through an inflection point at $\tilde{c}$, eventually reaching values as low as less than 1 at the highest investigated $c$. The number of anions entering the sphere and substituting to the solvent molecules increases in a complementary fashion, perfectly balancing at $c\simeq$~3~M the number of $\ce{EC}$ molecules (2). In the entire $c$-range, the number of co-ions included in $\mathcal{C}$ is negligible (diamonds), due to the strong repulsive Coulomb interactions.

The variation of the number of solvent molecules (and, consequently, anions) comprised in the coordination shell at different $c$ determines the average effective charge carried by the cation, $Q_C=\sum_{l\in\mathcal{C}} q_l$, whereas the thermal fluctuations of the chemical composition of $\mathcal{C}$ control the charge fluctuations $\delta Q_C$.
%(not shown~\cite{skarmoutsos2025}). 
In the inset of Fig.~\ref{fig:coordination_Li} we plot the $c$-dependence of $Q_\mathcal{C}$, directly determined from the coordination numbers data. For $c<c_o$ we have $Q_\mathcal{C}=1$, as expected given that $\ce{Li+}$ is the only species carrying a non-zero charge. At higher values of $c$ the charge steadily decreases, mirroring  the evolution of $n_{\ce{EC}}$ and changing sign at $c^*$. Therefore, for $c>c^*$ the dressed ion has a {\em negative} charge, which is as large as $-3e$ at the highest concentrations. \mymodA{(Note that the term "dressed" is reminiscent of the extensive work by R.~Kjellander~\cite{kjellander2016nonlocal,kjellander2018focus,kjellander2019intimate} but it does not have exactly the same meaning. In that theory the dressed charge corresponds to the charge carried by complexes of the size of a Debye-like length, determined by exploiting the non-locality of the electrostatics in terms of an eﬀective dielectric permittivity and a renormalized charge. In our analysis we did not attempt to employ that extremely sophisticated formalism, but kept an analogous physical content.)}

{\bf The ionic clusters phase.} The strictly local point of view of Fig.~\ref{fig:coordination_Li} has an obvious limitation. Indeed, the above procedure is perfectly reasonable at very low concentrations where, when applied to both co-ions, amounts to the mapping of the electrolyte onto a plasma of charged dressed ions, preserving the total charge neutrality. It fails, however, at high $c$ where it cannot properly treat anions which could pertain simultaneously to a few different $\mathcal{C}$, now leading to violation of the global charge neutrality. We can overcome this difficulty by relaxing the spherically symmetric view of the $\mathcal{C}$, and admitting the formation of extended ionic aggregates of any shape. This is a crucial point for the analysis below.

\mymodB{We have therefore performed a cluster formation analysis where ions are associated to ionic clusters on the basis of their mutual distances~\cite{allen2017computer}, such that each ion pertains to one and only one cluster. More precisely, a cluster is defined as a set of atoms, each of which is within the cutoff distance from one or more other atoms in the cluster. This procedure gives us access to all statistical features of ionic clusters. At each $c$, we have identified all clusters and generated the cluster size distribution for both solvent molecules and ions ($\ce{Li+}$ and $\ce{BF4-}$), \mymodB{using the cutoff values  $r_1=2.6$~\AA\; and $r_2=3.8$~\AA, respectively, determined as discussed above}. We show general features of the obtained distributions in Figs.~S5 to S7 in the supplementary material.} 

\mymodB{These distributions (Fig.~S5 a) in the supplementary material) are in general power laws with a cutoff at large cluster sizes, some of them presenting a peak at very high connectivity reminiscent of the formation of a spanning ionic cluster (inset of Fig.~S5 a) in the supplementary material, and see below). The statistics of the ionic clusters, provides us with important information, including the distribution probability of the clusters total charge, $Q_{\text{cl}}(c)$ (Fig.~S5 b) in the supplementary material), that we will employ in what follows. \mymodB{Although in this context the relevance of both clusters formation~\cite{mceldrew2020theory,mceldrew2021correlated,mceldrew2021ion,mceldrew2021salt,krucker2021underscreening,hartel2023anomalous,safran2023scaling} and the impact of complex ionic aggregates on the transport properties~\cite{france2019correlations,feng2019free} have already been recognized, we put them in a new perspective.}}
\begin{figure}[t]
\centering
\includegraphics[width=0.49\textwidth]{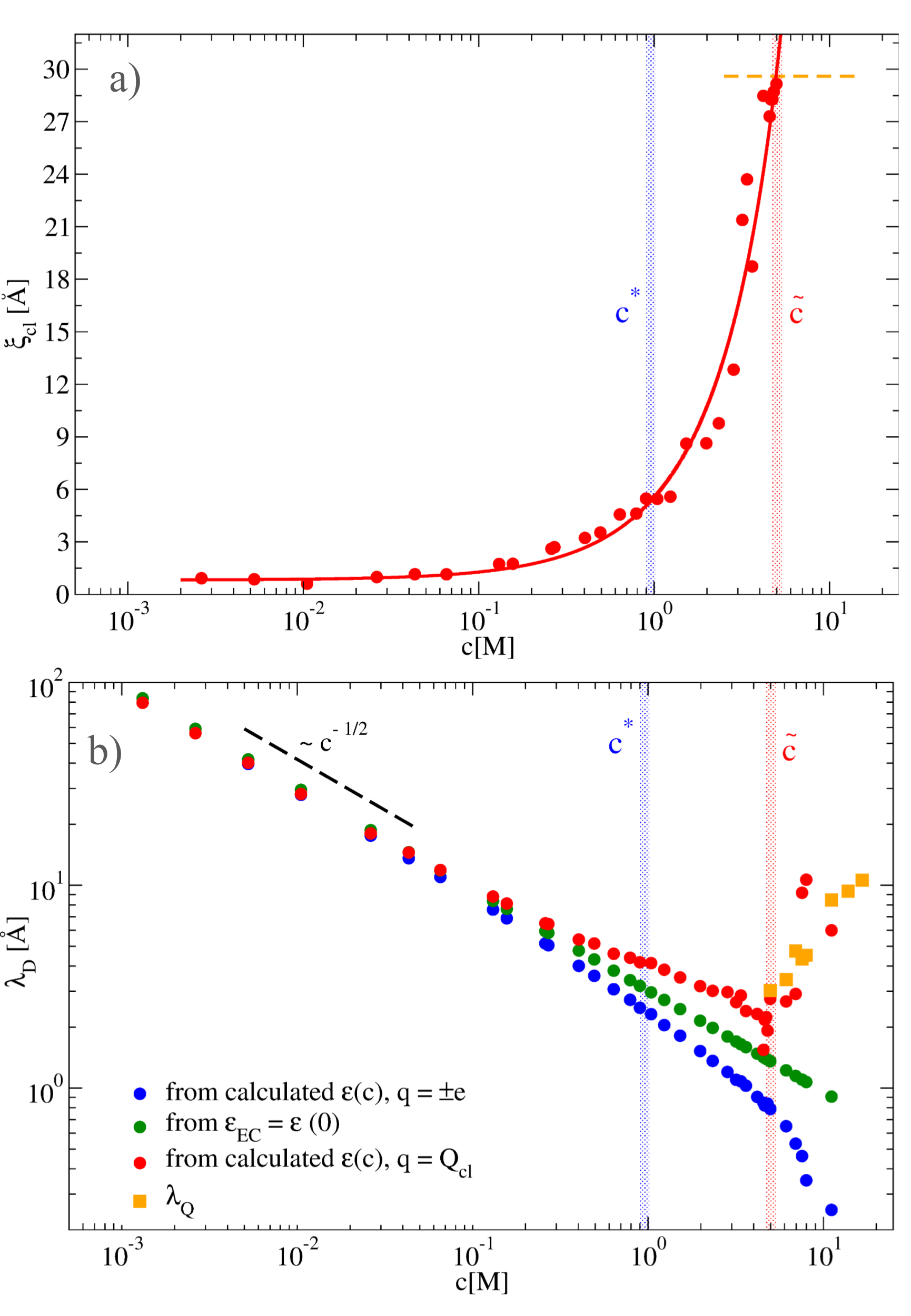}
\caption{Electrolyte length scales.  \mymodD{{\em a)} Length scale associated to the growing ionic domains (red circles), determined from Eq.~(\ref{eq:ionic_length}) as discussed in the text. We can follow the growth of $\xi_{\text{cl}}$ until reaching $\tilde{c}$, where $\xi_{\text{cl}}\simeq L/2\simeq$~30~\AA, (horizontal dashed line), which signals the formation of a fully-percolated cluster.} {\em b)} Debye-like length scales (circles), calculated composing different measured quantities, as discussed at length in the main text. All scales scale as $\sqrt{c}$ at low $c$ by construction, $\lambda_{cl}$ also grasps the high $c$ regular underscreening, superimposing to the $\lambda_Q$ data (squares) of Fig.~\ref{fig:modulus_and_s_q}~d). The vertical shaded lines indicate the position of $\tilde{c}$ (blue) and $c^*$ (red), respectively.}
\label{fig:final_lengths}
\end{figure}

{\bf Length scales.} Equipped with the above results, we are now in the position to identify and explicitly determine pertinent length scales that could be involved in the "mystery"~\cite{Holm2022} of the anomalous underscreening. By means of the coordinates $\left\{\mathbf{r}_i\right\}$ of the ions pertaining to clusters of size $l$, we obtain the typical average size of the clusters as~\cite{stauffer2018introduction},  
\begin{equation}
\xi_{\text{cl}}=\left[\frac{2\sum_l R_l^2l^2n(l)}{\sum_l l^2 n(l)}\right]^{1/2},
\label{eq:ionic_length}
\end{equation}
where $n(l)$ is the number of clusters of size $l$, $\sum_l n(l)l=N_+ + N_-$, and $R_l=[(1/2l^2)\sum_{i,j}|\mathbf{r}_i-\mathbf{r}_j|^2]$ is the gyration radius of an ionic cluster of size $l$. (Note that, in principle, other definitions of $\xi_{cl}$ are possible which, however, do not imply any qualitative difference, as verified in many contexts~\cite{stauffer2018introduction}). We show our data in Fig.~\ref{fig:final_lengths}~a). 
\begin{figure*}[t]
\centering
\includegraphics[width=1\textwidth]{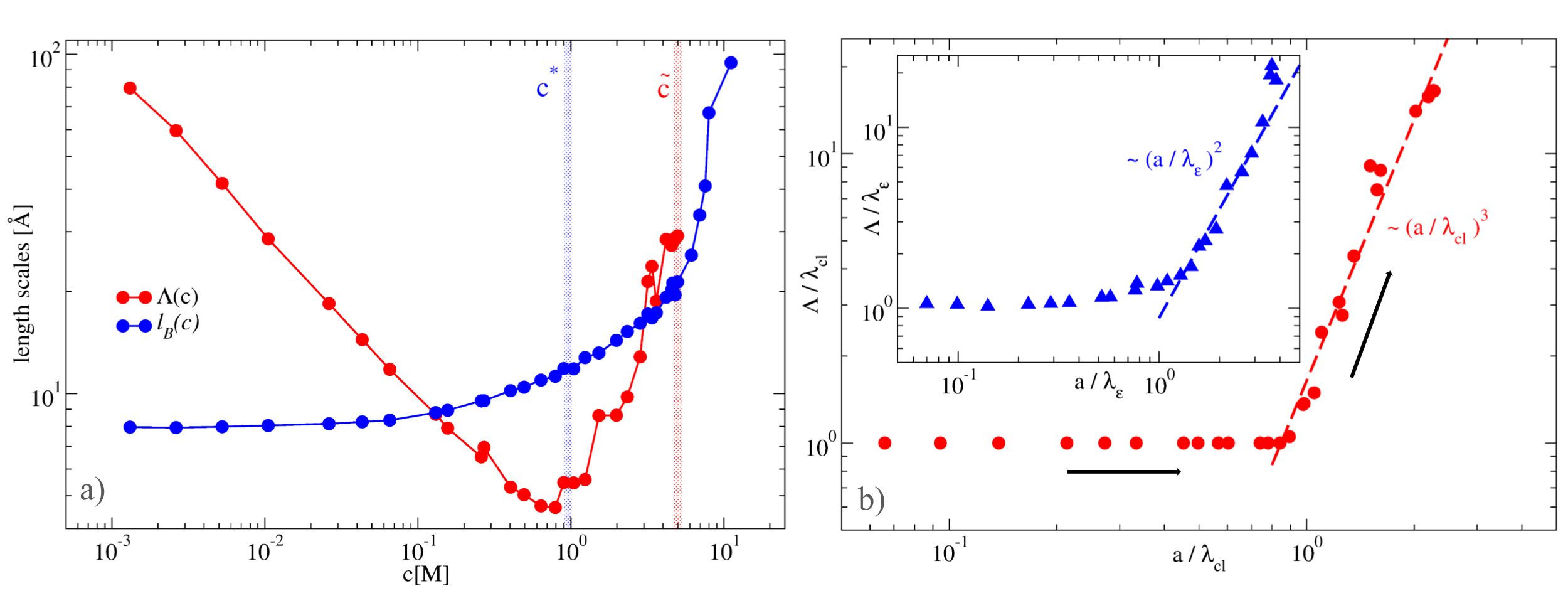}
\caption{Pertinent length scales and scaling law. {\em a)} We show with the red circles the length scale $\Lambda(c)$ defined by Eq.~(\ref{eq:my_lambda}) and the Bjerrum length, Eq.~(\ref{eq:bjerrum}) (blue circles). These data are discussed in details in the main text. {\em b)}. Scaling of the $\Lambda(c)$ data based on Eq.~(\ref{eq:scaling}), with $\lambda_D$ replaced by $\lambda_\epsilon$ (inset) and $\lambda_{cl}$ (main panel), respectively. The power laws with $\gamma=$~ 2 and 3 are also indicated with the dashed lines. The arrows indicate the increase of concentration following the parametric plot. The data are discussed in depth in the main text.}
\label{fig:final}
\end{figure*}

For $c<c_o$ we find $\xi_{\text{cl}}\simeq$~1\AA, which accounts for the average size of the few (rare) short-lived ion pairs which appear even at very low concentration. Interestingly, at $c\simeq c_o$, $\xi_{\text{cl}}$ starts to increase continuously with $c$, in a manner compatible with the critical behavior $|c-c_{\text{cr}}|^{-\nu}$, with $\nu=0.88$ and $c_{cr}=20.8$~M, analogous to the bulk modulus. These two evidences convincingly point toward the activation of a percolation process which determines the ionic domains growth on length scales continuously increasing with $c$, ultimately controlling the mechanical response modifications of the material. \mymodD{We can follow the growth of $\xi_{\text{cl}}$ until reaching $\tilde{c}$, where $\xi_{\text{cl}}\simeq L/2\simeq$~30~\AA, i.e., half of the size of the simulation box length at those concentrations (horizontal dashed line in Fig.~\ref{fig:final_lengths}~a)), which signals the formation of a fully-percolated cluster.}

We can make an additional step forward by noticing that, once determined the dielectric permittivity, we can directly build a few effective {\em Debye-like} length scales based on Eq.~(\ref{eq:debye_length}), by employing different options for a renormalized ionic charge, $Q^*$. A similar "constructive" methodology has been followed for the sake of the discussion in~\cite{gebbie2017long}, \mymodA{inspired by the dressed-ion concept by Kjellander~\cite{kjellander2016nonlocal,kjellander2018focus,kjellander2019intimate}}. We start by simply determining, as a reference, $\lambda_{\epsilon_{EC}}$ via Eq.~(\ref{eq:debye_length}) with $Q^*=e$ and a $c$-independent $\epsilon=\epsilon_{EC}$, with $\epsilon_{EC}$ the calculated value pertaining to the pure \ce{EC}, in the considered thermodynamic conditions. We show the data in Fig.~\ref{fig:final_lengths}~b), in green.  $\lambda_\epsilon$ determined  with the calculated $\epsilon(c)$ (blue), already gives a quite different picture, with the $\propto\sqrt{c}$ dependence at low-$c$ followed by a faster decay and a sharp cutoff around $c=\tilde{c}$.

We now build an additional length scale by composing the calculated dielectric permittivity data of Fig.~\ref{fig:epsilon_and_sigma}~a) with the distribution of the clusters charge $Q_{cl}$ discussed above as, 
\begin{equation}
\lambda_{\text{cl}}=\left [\frac{k_B T V\epsilon_o\epsilon}{\sum_{\textit{cl}} Q_{\textit{cl}}^2}\right]^{1/2}.
\label{eq:debye_lengths}
\end{equation}
Interestingly, while $\lambda_{\text{cl}}$ (red circles) decreases at low and intermediate $c$, as expected, it suddenly starts to increases at $\tilde{c}$. Most surprising, we also plot (yellow squares) the $\lambda_Q(c)$ we estimated by fitting the spatial-charge distribution function $Q(r)$ discussed above (Fig.~\ref{fig:modulus_and_s_q}~d)), which matches the $\lambda_{\text{cl}}$. 

This finding indicates that $\lambda_{\text{cl}}$ not only grasps (by construction) the electrolyte physics in the diluted limit, but it also captures the expected regular underscreening at high concentrations. We therefore keep $\lambda_{\text{cl}}$ as the most appropriate length scale associated to screening. We stress at this point, that no adjustable parameter is involved in these results, which we simply determine by employing independently calculated system properties together with a re-normalized charge, finally only based on physical soundness and dimensional arguments. 

{\bf Anomalous underscreening: a proposal.} We have been able to identify two length scales following opposite trends with concentration. On one side, $\lambda_{\text{cl}}$ (red circles in Fig.~\ref{fig:final_lengths}~b)) decreases with $c$ from a magnitude as high as 10~nm to values shorter than the mean ionic size. On the other side, $\xi_{\text{cl}}$ (Fig.~\ref{fig:final_lengths}~a)) continuously increases in the entire $c$-range, reaching values only limited by the finite size of the simulation box. Note that these scales quantify different physical parameters: while the former accounts for the decreasing range of the screened interactions between the renormalized charged entities (ionic clusters), the latter quantifies the increasing spatial extent of the same entities. 

Guided by the idea that the {\em largest} of these lengths should be the most relevant in the different $c$-conditions, we build an additional length, 
\begin{equation}
\Lambda(c)=\max{\{\lambda_{cl}(c), \xi_{cl}(c)\}}.
\label{eq:my_lambda}
\end{equation}
This is the most important point of the paper, and we show the data in Fig.~\ref{fig:final}~a) (red circles). In this representation, $\Lambda(c)$ is very similar to the findings of the experiments with the SFB, as reported in~\cite{smith2016electrostatic} and others, with the expected decreasing branch at low concentrations, followed by the strong growth and intermediate-to-high $c$ associated to the anomalous underscreening. $\Lambda$ reaches the minimum value, $\Lambda_{\text{min}}\simeq$~4.5~\AA, at the concentration where $Q(r)$ of Fig.~\ref{fig:modulus_and_s_q}~c) transition to the oscillatory exponential decay, i. e., at the Kirkwood point. \mymodB{Note that $\Lambda_{\text{min}}$ corresponds to the mean ion diameter, $a\simeq$~4.5\,\AA\,$\simeq 2\pi/q_{\text{max}}$, where $q_{\text{max}}\simeq$~1.4~\AA$^{-1}$ is the position of the first diffraction peak of the calculated partial static structure factor, $S_{\ce{Li}-\ce{BF4}}(q)$ at $c^*$. (To the best of our knowledge, this is the first time that a length scale increasing with $c$, and substantially larger than those associated to regular underscreening, is convincingly detected in a simulated electrolyte.)} 

In Fig.~\ref{fig:final}~a) we also show the Bjerrum length, determined from the calculated $\epsilon(c)$ as,
\begin{equation}
l_B=\frac{e^2}{4\pi \epsilon_o\epsilon
k_B T},
\label{eq:bjerrum}
\end{equation}
which quantifies the strength of the electrostatic interactions and determines the scale below which direct electrostatic interactions dominate over thermal fluctuations~\cite{bocquet2010nanofluidics}. Interestingly, we find $\Lambda<l_B$ in the concentration interval where the difference between the true calculated ionic conductivity and $\sigma_{\text{NE}}$ is maximum (Fig.~\ref{fig:epsilon_and_sigma}~b)). The detrimental contribution of the collective motions of the ions to transport is therefore consistently maximum corresponding to the maximum strength of the electrostatic interactions, highlighting the interplay between screening features and transport properties.  
%Second, we discuss a last point that additionally strengthens the soundness of our interpretation of the data. 

\mymodC{We conclude by addressing a last issue. It has been demonstrated~\cite{lee2017scaling,lee2017underscreening} that for a quite large range of systems with different ions chemistry (including inorganic salts in water, pure ionic liquids, mixtures of ionic liquids with organic solvents) the scaling relation exists,
\begin{equation}
\frac{\lambda_S}{\lambda_D}\propto
\begin{cases}
1                                 & a / \lambda_D \ll 1 \\
\left(a / \lambda_D\right)^\gamma &  a / \lambda_D \gtrsim 1,
\end{cases}
\label{eq:scaling}
\end{equation}
which collapses all experimentally measured screening lengths, $\lambda_S$, on a single "master curve". Here, the exponent $\gamma=$~3 when the anomalous underscreening is involved, $a$ is the mean ionic radius, and the apparent $\lambda_D$ is calculated from Eq.~(\ref{eq:debye_length}) in the entire considered concentration range employing the measured dielectric constants. It is tempting to test our data against the above scaling.} 

\mymodC{In the inset of Fig.~\ref{fig:final}~b) we show the parametric representation of Eq.~(\ref{eq:scaling}) with $\lambda_S=\Lambda$ and $\lambda_D=\lambda_\epsilon$ (blue symbols of Fig.\ref{fig:final_lengths}), on a double-logarithmic scale. The data qualitatively fit the predicted picture, with the expected unit value assumed below the Kirkwood point, followed by a clear power-law dependence (blue dashed line) maintained up to $\tilde{c}$. In contrast, we find $\gamma=2$ for the exponent, which is at variance with the experiment.}

\mymodC{A different conclusion, however,  can be drawn by recalling that in our simulations $\lambda_{cl}$ (determined from the {\em ionic clusters charges}) turns out to be more appropriate than $\lambda_\epsilon$ (determined from the {\em bare ions charges}) in describing the electrolyte screening features at moderate to high concentrations. As a consequence, it does make sense to employ $\lambda_{cl}$ in the same parametric representation, that we show in the main panel of Fig.~\ref{fig:final}~b).}

\mymodC{Intriguingly, following this route we recover the $\gamma=$~3 scaling, in a picture which now naturally involves ionic-clusters-related length scales only, substantially deviating from the analysis of the experimental data on one hand, but apparently grasping the same phenomenology, on the other. Note that in ~\cite{lee2017scaling} a detailed derivation of the cubic scaling is provided involving charge {\em defects}, which, despite their different nature, appear to play a role analogous to the renormalized charges of the ionic clusters discussed here. Work is in progress to further clarify this matter.} 

\section*{Conclusions}
\label{sec:conclusions}
Through extensive Molecular Dynamics simulations, we have thoroughly investigated the properties of a generic model electrolyte, $\ce{Li+}\ce{BF4^-}$ dissolved in $\ce{EC}$, across a wide range of salt concentrations. This spectrum consistently covers any allowed state in the considered external thermodynamic conditions, ranging from the ideal non-interacting Debye limit, through intermediate and high concentrated mixtures, to solvent-in-salt system instances. We have precisely quantified structural, dielectric, and transport changes induced by varying concentrations, and elucidated the resulting nanoscale organization of the ions. 

Most important, on one side, \mymodA{inspired by the dressed ion view applied to charged extended ionic domains instead of coordination spheres of size comparable to the Debye length,} we have been able to identify the length scale associated to the screening features of these domains. On the other side, we have directly quantified the extent of such domains, highlighting the existence of a percolation process, eventually terminating in the formation of a gel formed by ionic clusters. We have demonstrated that the outcomes of SFB experiments find a natural description in this view, at the expenses of a rather different interpretation of the data. While the detected anomalous increasing length scale indeed turns out to be an electrolyte bulk feature, it does not relate to the range of the electrostatic interaction (pair property) but rather to the extent of the growing ionic domains (collective property). \mymodB{Note that the relevance of clustering processes in the present matters have been often advocated in the past, here we integrate them in a comprehensive natural picture.}

We conclude with a few observations. First, we acknowledge that the bulk ionic fluid layout we consider in our work is significantly different than the experimental configuration, where the electrolyte is confined between two surfaces. The {\em experimentum crucis} unambiguously connecting the simulation approach to the experimental setup would imply to directly measure a long-range exponential tail in the force profile between two surfaces immersed in the electrolyte, similar to~\cite{pivnic2019structural}, while {\em simultaneously} detecting the growth of the ionic domains we suggested here. 
%\mymodC{Add here an observation about the behavior at temperatures where solid phases are present in particular concentration ranges?.} Work is in progress in this direction. 

\mymodB{Next, we note that the point of view we have employed for the analysis of the ionic domains is intrinsically static, although our discussion of the interplay between $\Lambda$ and the Bjerrum length seems to indicate that this approach already bears non-trivial information on transport features. It would still be interesting, however, to clarify how finite life-times of these charged structures contribute to determining the electrolyte transport properties, similar to~\cite{feng2019free,jones2021bayesian,elliott2024dynamic}.} 

Also, our results could allow new insight on the development of Casimir forces in dense confined electrolytes as discussed~\cite{lee2018casimir}, by connecting them to the observed percolation process, as suggested in~\cite{gnan2014casimir}. Finally, it will be appealing to explore the implications of our findings on the features of the electric double layer capacitance, important in the design and optimization of devices like super-capacitors when seeking for high energy storage performances.

\section*{Methods} 
\label{sec:methods}
Numerical methods employed for the Molecular Dynamics simulations and the subsequent data analysis are standard. We give below a few details, while more extended information \mymodB{can can be found in the supplementary material}, and in our previous~\cite{skarmoutsos2015li+}. 

\mymodB{We have considered an all-atoms description for a ternary mixture, formed by lithium tetrafluoroborate, \ce{Li+BF4-}, dissolved in ethylene-carbonate, \ce{EC}. We have employed the inter-molecular force field of~\cite{soetens1998molecular,masia2004ethylene,skarmoutsos2015li+} (we report all force-field parameters in Tab.~S1 in the supplementary material) with the difference that, while keeping the molecular structures of~\cite{soetens1998molecular}, we have disregarded the intramolecular interactions considering the molecules as rigid. We do not expect any significant qualitative difference induced by this modifications on our long-time dynamical results, nor particular improvements coming from more recent developments, like the Machine Learning force field of~\cite{magduau2023machine}. Changes, in contrast, could be induced by a non-polarizable force field, that would modify the behavior of the total dipole associated to the cation coordination sphere, a crucial feature as discussed at length in~\cite{skarmoutsos2015li+}. This is an acceptable limitation in this work, where it is crucial to count on simple modeling of very large systems on very long time scales. \mymodB{We observe, however, that as pointed out above, the present force field grasps extremely efficiently the physics of the investigated system at low concentrations, including dielectric and ionic transport properties, as demonstrated by the comparison with the available experimental data.}}

The time evolution has been generated with the Large-scale Atomic Molecular Massively Parallel Simulator (LAMMPS)~\cite{LAMMPS}. The long-range part of the Coulomb interactions have been determined via the particle-particle particle-mesh (PPPM) method, with a desired relative error in forces $\delta F=10^{-5}$. Equations of motions have been integrated numerically with a time step $\delta t=$~1~fs, separately for the point-like Lithium ion, and the rigid \ce{CE} and \ce{BF_4^-} molecules. For the former we have employed the Nose-Hoover non-Hamiltonian equations of motion~\cite{allen2017computer,tuckerman2023statistical}, designed to generate positions and velocities sampled from the canonical (NVT) ensemble. For the latter, the equations of motions were integrated in the NPT ensemble, using a Nose/Hoover barostat with chains~\cite{allen2017computer,tuckerman2023statistical}. At each time step the total force and torque on each rigid body was computed as the sum of the forces and torques on its constituent particles. The coordinates, velocities, and orientations of the atoms in each body where then updated so that the body moved and rotated as a single entity. \mymodB{We have checked that this techniques allows to correctly relax the box size to match the electrolyte equilibrium density at the fixed $(P, T)$ (see Fig.~S1 in the supplementary material) and data in Tabs.~S2 and~S3 in the supplementary material, and verified that both the salt and solvent components kinetic energies conforming to the chosen value of $T$.}

We have considered a total of 40 independent systems, corresponding to an extremely wide range of concentrations \mymodA{(molar concentrations in mol/L)} (see Tabs. SII and SIII in the supplementary material), at temperature $T=$~400~K and ambient pressure. Each system has been simulated for 30~ns, following much longer thermalization runs to ensure proper thermodynamic equilibrium, as already mentioned. A series of out-of-equilibrium runs under the effect of an external electric field were also performed to determine both the dielectric permittivity and the ionic conductivity via the finite field formalism of~\cite{cox2019finite}, as already detailed above.
\section*{Supplementary material}
See the supplementary material for the details of the simulations and additional data mentioned in the main text.
\section*{Acknowledgments}
This work was supported by the French National Research Agency under the France 2030 program (Grant ANR-22-PEBA-0002). S. M. also acknowledges support by the project MoveYourIon (ANR-19-CE06/0025) funded by the French "Agence Nationale de la Recherche".
\section*{Author Declarations}
\subsection*{Conflict of interest}
The authors have no conflicts to disclose.
\section*{Data availability}
The data that support the findings of this study are available from the corresponding author upon reasonable request.
\bibliography{references}
\end{document}